\documentclass[sigconf]{acmart}
\usepackage{booktabs} 
\usepackage{subfigure}
\usepackage{amsmath}
\usepackage{float}
\usepackage{graphicx}
\usepackage{amsfonts}
\usepackage{amsthm}
\usepackage{dblfloatfix}
\usepackage{marvosym}
\usepackage[ruled,vlined]{algorithm2e}
\newtheorem{problem}{Problem}
\makeatletter
\newcommand{\removelatexerror}{\let\@latex@error\@gobble}
\makeatother
\setcopyright{acmcopyright}

\acmDOI{xx.xxx/xxx_x}

\acmISBN{978-1-4503-8713-2/22/04}

\copyrightyear{2021}
\acmYear{2021}
\setcopyright{acmlicensed}\acmConference[WI-IAT '21 Companion]{IEEE/WIC/ACM International Conference on Web Intelligence}{December 14--17, 2021}{ESSENDON, VIC, Australia}
\acmBooktitle{IEEE/WIC/ACM International Conference on Web Intelligence (WI-IAT '21 Companion), December 14--17, 2021, ESSENDON, VIC, Australia}
\acmPrice{15.00}
\acmDOI{10.1145/3498851.3498926}
\acmISBN{978-1-4503-9187-0/21/12}

\begin{document}
\title{Heterogeneous Graph Learning for Explainable Recommendation over Academic Networks}
  
\renewcommand{\shorttitle}{Heterogeneous graph Attention InfoMax}

\author{Xiangtai Chen}
\orcid{0000-0002-6619-9396}
\affiliation{%
  \department{School of Software}
  \institution{Dalian University of Technology}
  \streetaddress{}
  \city{Dalian} 
  \country{China} 
  \postcode{}
}
\email{chenxiangtai@outlook.com}

\author{Tao Tang}
\affiliation{%
  \department{School of Engineering, IT and Physical Sciences}
  \institution{Federation University Australia}
  \streetaddress{}
  \city{Ballarat} 
  \country{Australia} 
  \postcode{}
}
\email{tau.tang@outlook.com}

\author{Jing Ren}
\affiliation{%
  \department{School of Engineering, IT and Physical Sciences}
  \institution{Federation University Australia}
  \streetaddress{}
  \city{Ballarat} 
  \country{Australia} 
  \postcode{}
}
\email{ch.yum@outlook.com}

\author{Ivan Lee\textsuperscript{\Letter}}
\affiliation{%
  \department{STEM}
  \institution{University of South Australia}
  \city{Adelaide}
  \country{Australia}
}
\email{ivan.lee@unisa.edu.au}

\author{Honglong Chen}
\affiliation{%
  \department{College of Control Science and Engineering}
  \institution{China University of Petroleum}
  \city{Qingdao}
  \country{China}
  \postcode{266580}
}
\email{chenhl@upc.edu.cn}

\author{Feng Xia}
\affiliation{%
  \department{School of Engineering, IT and Physical Sciences}
  \institution{Federation University Australia}
  \streetaddress{}
  \city{Ballarat} 
  \country{Australia} 
  \postcode{}
}
\email{f.xia@ieee.org}

\renewcommand{\shortauthors}{Xiangtai Chen et al.}

\begin{abstract}
    With the explosive growth of new graduates with research degrees every year, unprecedented challenges arise for early-career researchers to find a job at a suitable institution. This study aims to understand the behavior of academic job transition and hence recommend suitable institutions for PhD graduates. Specifically, we design a deep learning model to predict the career move of early-career researchers and provide suggestions. The design is built on top of scholarly/academic networks, which contains abundant information about scientific collaboration among scholars and institutions. We construct a heterogeneous scholarly network to facilitate the exploring of the behavior of career moves and the recommendation of institutions for scholars. We devise an unsupervised learning model called HAI (Heterogeneous graph Attention InfoMax) which aggregates attention mechanism and mutual information for institution recommendation. Moreover, we propose scholar attention and meta-path attention to discover the hidden relationships between several meta-paths. With these mechanisms, HAI provides ordered recommendations with explainability. We evaluate HAI upon a real-world dataset against baseline methods. Experimental results verify the effectiveness and efficiency of our approach.
\end{abstract}

%
%



\maketitle

\section{Introduction}\label{sec:introduction}
Recent years have witnessed the rapid growth of academic information (i.e., big scholarly data) due to a large number of research works carried out by academia and industry~\cite{xia2017big}. Meanwhile, academic social networks are continuously expanded.
It is well recognized that different types of academic social relationship have intrinsically-different effect among researchers, which form a complex force that influence the dynamics of academic social networks. From the UNESCO Science Report\footnote{\url{https://en.unesco.org/node/252277}}, the number of full-time equivalent researchers grew by 21\% from 2007 to 2013, accounting for 0.1\% of the global population. Nowadays, the number of doctoral graduates is expanding every year, which results in input inflation of academic researchers. It is worse for early-career researchers that industry might be unable to absorb such numerous researchers, while academic institutions are generally adopting the Tenure-Track on recruiting researchers. With such fierce competition, doctoral graduates are facing a dilemma when choosing institutions.

However, a survey about PhD degrees in nature reveals that despite facing a love–hurt relationship, doctoral students are as committed as ever on pursuing research careers~\cite{woolston2017graduate}. The first choice of which academic institution to join has a great impact on the future academic career of doctoral graduates, further choosing a suitable institution could contribute to future academic success. But how can the doctoral students find a research position? The Nature survey~\cite{baker2015social} indicates that doctoral students are largely finding their career advice online. Just one-third credited advice from a supervisor as a reason for their career choice. With the rapid growth of social media, researchers can use such platforms liked Twitter to expand their social contacts and find jobs~\cite{baker2015social}.  Although some platforms such as LinkedIn aim to build a career network for job hunting, early-career researchers would like to find an academic position through friendship relationships and collaboration relationships on Twitter.

Utilizing social media to find an academic job is time-consuming and heavily relies on manually posting, which greatly limits its widespread usage. An ideal solution is to design a method to automatically discover the hidden collaboration information, colleague information, academic research direction, etc. from academic social networks (ASNs). Nevertheless, accurately distinguishing social relationships is difficult, especially in a real-world network. In the real data, the vast majority of the patterns of career moves are hidden in the collaboration information, which is not currently revealed in most social networks.

Currently, the problem of finding an optimal academic institution for early-career researchers is not well addressed, and existing social networks are not sufficient to address the issues about finding academic institutions on a large scale.

In this work, we aim to tackle the above problem. First, we construct an undirected heterogeneous scholarly network (HSN) with various types of nodes and edges, where nodes include scholars, institutions, papers, while edges include "works-with" and "writes". Furthermore, we design a model called HAI which works on automatically discovering collaboration information and colleague information for institution recommendation for junior scholars. We adopt attention mechanism on the meta-path neighbors connected by two meta-paths: Author-Paper-Author (\emph{APA}) and Author-Institution-Author (\emph{AIA}). The scholar mutual information is adopted to maximize the similarity of local  features  and  global  features. Generally, there has no label to indicate which academic institution the scholar will join in the future for a recommendation system. To address the problem of lack of labels in the real-world data, we maximize the scholar mutual information as the objective function and apply unsupervised learning to learn the information of scholars. Moreover, the experimental results show that our model performs well in terms of the AUC metric and HR metric.

The primary contributions of this work are as follows:
\begin{itemize}
    \item We devise a novel unsupervised learning algorithm to learn low-dimension features of scholars by taking advantage of the attention mechanism and mutual information of scholars.
    \item We proposed an explainable recommendation model (i.e., HAI) which can recommend suitable academic institutions for early-career researchers while providing explanations.
    \item Extensive experiments have been conducted upon a real-world dataset. The results verify the superiority of HAI against state-of-the-art baseline methods.
\end{itemize}

The rest of this paper is organized as follows: Section~\ref{section2} gives a discussion of the related works. Section~\ref{section3} presents the primary preliminaries of heterogeneous graph learning and defines the problem we are dealing with. Section~\ref{section4} proposes the method including scholar attention, meta-path attention, and scholar mutual information to recommend institutions for junior scholars. Section~\ref{section5} shows the data analysis, experimental results, and case studies. In the end, Section~\ref{section6} concludes this study and discusses the future directions.

\section{Related work}\label{section2}

Meta-path, as an important characteristic of a heterogeneous network, is regarded as a useful tool for heterogeneous network embedding~\cite{xia2021graph}. Many researchers design algorithms based on meta-path and heterogeneous neighborhoods generated by random walks~\cite{dong2017metapath2vec,fu2017hin2vec,xia2020random}. By exploring meta-paths, node-level attentions are present to learn heterogeneous relations~\cite{wang2019heterogeneous,lu2019relation}. Random walks on heterogeneous structures walk slowly on meta-path, while attention on HAI runs much faster than random walks. Furthermore, neural networks are proposed to embed heterogeneous networks. Generative or discriminative adversarial networks based framework~\cite{hu2019adversarial,qu2018curriculum} works on complex neural networks to learn the node distribution. Qu et al.~\cite{qu2018curriculum} study learning curricula for node embedding in heterogeneous star networks and propose an approach based on deep reinforcement learning for this problem. Zhang et al.~\cite{zhang2019heterogeneous} propose a heterogeneous graph neural network model which uses two modules to aggregate feature information of sampled neighboring nodes. Wang et al.~\cite{wang2021scholar2vec} represent scholars to vector for lifetime collaborator prediction.

Network embedding has shown great power in the analysis of homogeneous networks~\cite{hou2020network}. HAI is different from the existing studies in heterogeneous network embedding. The previous works mainly focus on meta-path embedding or employ attention mechanism on labeled data for supervised learning, while HAI aggregates attention mechanism and mutual information on HSN for unsupervised learning.

In the early age of recommendation systems, Collaborative Filtering~\cite{schafer2007collaborative} and Matrix Factorization are two of the most widely used algorithms in the industry. Wan et al.~\cite{wang2020collaborative} recommend citations with network representation learning by Collaborative Filtering and~\cite{wan2020deep} utilize deep Matrix Factorization for Trust-Aware recommendation in social networks. Zhao et al.~\cite{zhao2017meta} introduce Matrix Factorization and Factorization Machine to learn similarities generated by each meta-path for the recommendation.

Nowadays, various kinds of embedding information are integrated for item recommendation~\cite{shi2019heterogeneous}.
Meta-path contexts~\cite{hu2018leveraging} are leveraged for TOP $K$ recommendation.
Wang et al.~\cite{wang2019unified} assume there exist some common characteristics under different meta-paths for each user or item and propose a unified embedding model.
Zhao et al.~\cite{zhao2020hetnerec} construct a heterogeneous co-occurrence network in a recommendation-oriented heterogeneous network.
Ying et al.~\cite{ying2018graph} develop a data-efficient graph convolutional network for web-scale recommender systems.

More recently, embedding information is applied to mine the relationships hidden in ASNs. In~\cite{wang2020attributed}, academic collaboration networks are built for academic relationship mining. ~\cite{ren2019api} and ~\cite{wiechetek2020analytical} design metrics to evaluate the academic potentials of scholars and academia units respectively. Guo et al.~\cite{guo2020graduate} make predictions on graduate employment with bias, but not providing suggestions for graduate employment. In ~\cite{wang2021attractive}, attractive communities are detected in ASNs and Yu et. al~\cite{yu2021how} optimize academic teams when the outlier member is leaving.
However, none of the previous works focus on helping scholars to deal with the career problem. For tackling the problem, we design a novel model HAI for academic institution recommendation.

\section{Preliminaries}\label{section3}
In this section, we introduce the input to our framework. The input HSN is a kind of Heterogeneous Information Network (HIN) with various types of nodes and various types of edges and has much more complicated structural information than homogeneous networks.
\subsection{Heterogeneous Graph}
\begin{definition}[Heterogeneous Graph]
    The input graph can be denoted as $\mathcal{G}=(\mathcal{V} ,\mathcal{E},\mathcal{A} ,\mathcal{X}  )  $, where $\mathcal{V} $ denotes a set of nodes with different types, $\mathcal{E} $ denotes a set of edges with different types, $\mathcal{A}$ denotes the set of adjacency matrices for each type of edge, and $\mathcal{X}$ denotes the set of matrices of nodes features for each type of node.
\end{definition}

Each node $v\in \mathcal{V} $ and each edge $e\in \mathcal{E} $ are associated with type mapping functions $\tau(v):\mathcal{V}\rightarrow \mathcal{T} ^v  $ and $\phi (e):\mathcal{E}\rightarrow \mathcal{T} ^e$, where $\mathcal{T} ^v$ and $\mathcal{T} ^e $ represent the set of node types and edge types respectively.
The graph becomes homogeneous when $ \vert\mathcal{T} ^v\vert = \vert\mathcal{T} ^e\vert = 1$, in contrast the graph is considered as heterogeneous when $ \vert\mathcal{T} ^v\vert + \vert\mathcal{T} ^e\vert> 1 $.
The whole nodes' information in the graph is stored in a set of node feature matrices $\mathcal{X} =\{ \mathbf{X} _1,\mathbf{X} _2,\ldots \mathbf{X} _n \}  $, where $n=\vert\mathcal{T} ^v\vert$; $\mathbf{X} _n\in \mathbb{R} ^    {N_n\times D_n}$ is the feature matrix of all nodes with type $\mathcal{T} ^v_n$; $N_n$ is the number of nodes in type $\mathcal{T} ^v_n$; and $D_n$ is the dimension of each node's embedding.
The edge information is present as a set of adjacency matrices $\mathcal{A} =\{ \mathbf{A} _1,\mathbf{A} _2,\ldots \mathbf{A} _m\}  $, where $m=\vert\mathcal{T} ^e\vert$.

\subsection{Meta-Path}
A remarkable difference between heterogeneous network and homogeneous network is the connection information among various types of nodes.
The edge $e=(s,d)$ connected from source node $s$ to destination node $d$ of a different type is denoted as $\langle\tau(s),\phi(e),\tau(d) \rangle $, and the edge is considered as meta-relation.
\begin{definition}[Meta-Path]
    Meta-path denoted as $\rho $ is a sequence of those meta-relations, which can be defined in the form of sequence $v_1\stackrel{\phi(e_1)}{\longrightarrow} v_2\stackrel{\phi(e_2)}{\longrightarrow}\ldots \stackrel{\phi(e_l)}{\longrightarrow}v_{l+1}$ connected by $l$ meta-relations.
\end{definition}

\subsection{Meta-Path Neighbors}
\begin{definition}[Meta-Path Neighbors]
    For a node $v$ and meta-path $\rho $ in heterogeneous network, the nodes which linked with node $v$ through meta-path $\rho $ are defined as meta-path neighbors denoted by $\mathcal{N}^\rho_v $.
\end{definition}
 The node $v$ itself is also included in the set of meta-path neighbors because of attention operations.

\begin{figure}[h]
    \centering
    \includegraphics[width=3.0in]{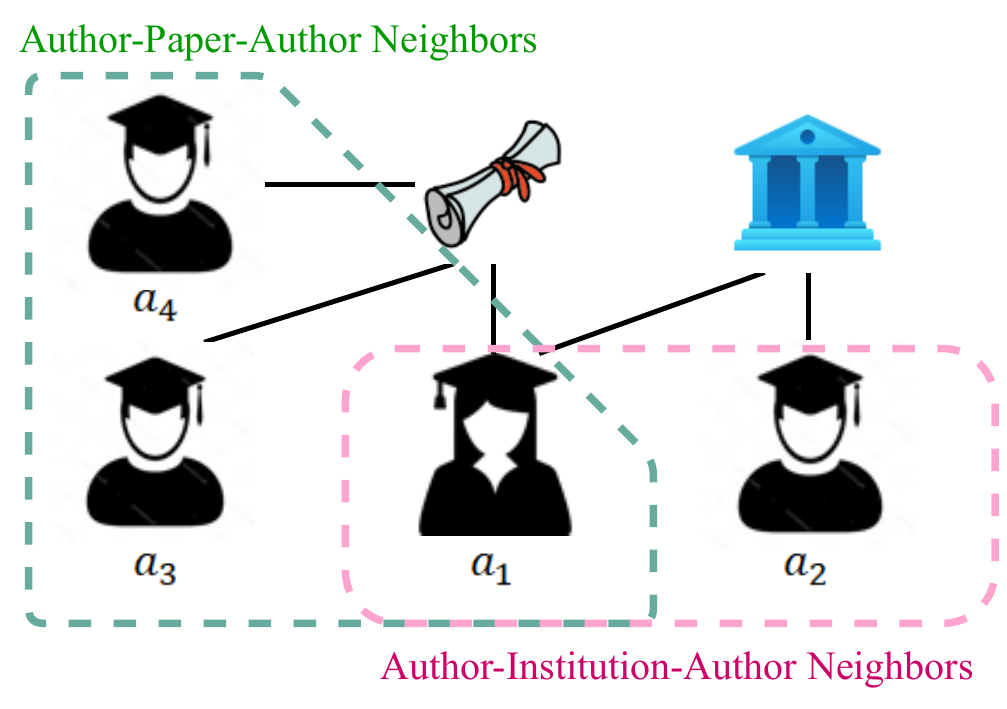}
    \caption{An instance of meta-path neighbors. The scholars in the area bounded by the green dotted line are Author-Paper-Author neighbors of $a_1$. The scholars in the area bounded by the red dotted line are Author-Institution-Author neighbors of $a_1$.}
    \label{metapath}
\end{figure}

\begin{example}
    As shown in Figure~\ref{metapath}, the meta-path neighbors of $a_1$  through meta-path Author-Paper-Author include $a_3$, $a_4$, and $a_1$ itself. Those meta-path neighbors on meta-path Author-Paper-Author are authors that $a_1$ has collaborated with. Similarly, the meta-path neighbors of $a_1$ through meta-path Author-Institution-Author contain the authors that work with the same institution of $a_1$, which are $a_2$, $a_1$ from Figure~\ref{metapath}.
\end{example}

\subsection{Problem Formulation}
The proposed model HAI is denoted as function $\mu $. Each scholar has an ordered preference list on institutions in the form of $Pre(a) = (i_1,i_2,\ldots i_s )$, where $s=\vert\mathcal{V}^i\vert$.
For a specific scholar $a_i\in\mathcal{V}^a$, we generate a preference list $Pre(a_i)=(i_1,i_2,i_3,i_4\ldots)$ for $a_i$ in $\mathcal{V}^a$. Assuming the institution $i_3$ is where the scholar $a_i$ works with, the preference list $Pre(a_i)$ indicates that $a_i$ may prefer $i_1$ to $i_2$ and prefers staying in $i_3$ instead of moving to $i_4$. The problem we addressed is to recommend institutions $\mathcal{I}_i $ for the scholar $a_i$ from the preference list $Pre(a_i)$.

The institution recommendation problem in this research is defined as follows:

\begin{problem}
    Given a preference list $Pre(a)=\mu (a)\in \mathcal{V}^i $ for scholar $a$, the target is to recommend institutions $\mathcal{I}  \in Pre(a)$ for any $a\in\mathcal{V} ^a$.
\end{problem}

\section{Design of HAI}\label{section4}
In this section, we propose a novel unsupervised heterogeneous graph learning model HAI for institution recommendation. From the inspiration of HAN~\cite{wang2019heterogeneous}, the model we proposed utilizes scholar attention to learn the attention score of meta-path neighbors and meta-path attention to learn the node embeddings from the meta-paths. Despite HAN has good performance on node classification, it could not be used for the institution recommendation because HAN is a semi-supervised learning algorithm that needs labels. Hence, we import mutual information mechanism to fill this gap.

\subsection{Overall Framework}
\begin{figure}[h]
    \centering
    \includegraphics[width=\linewidth]{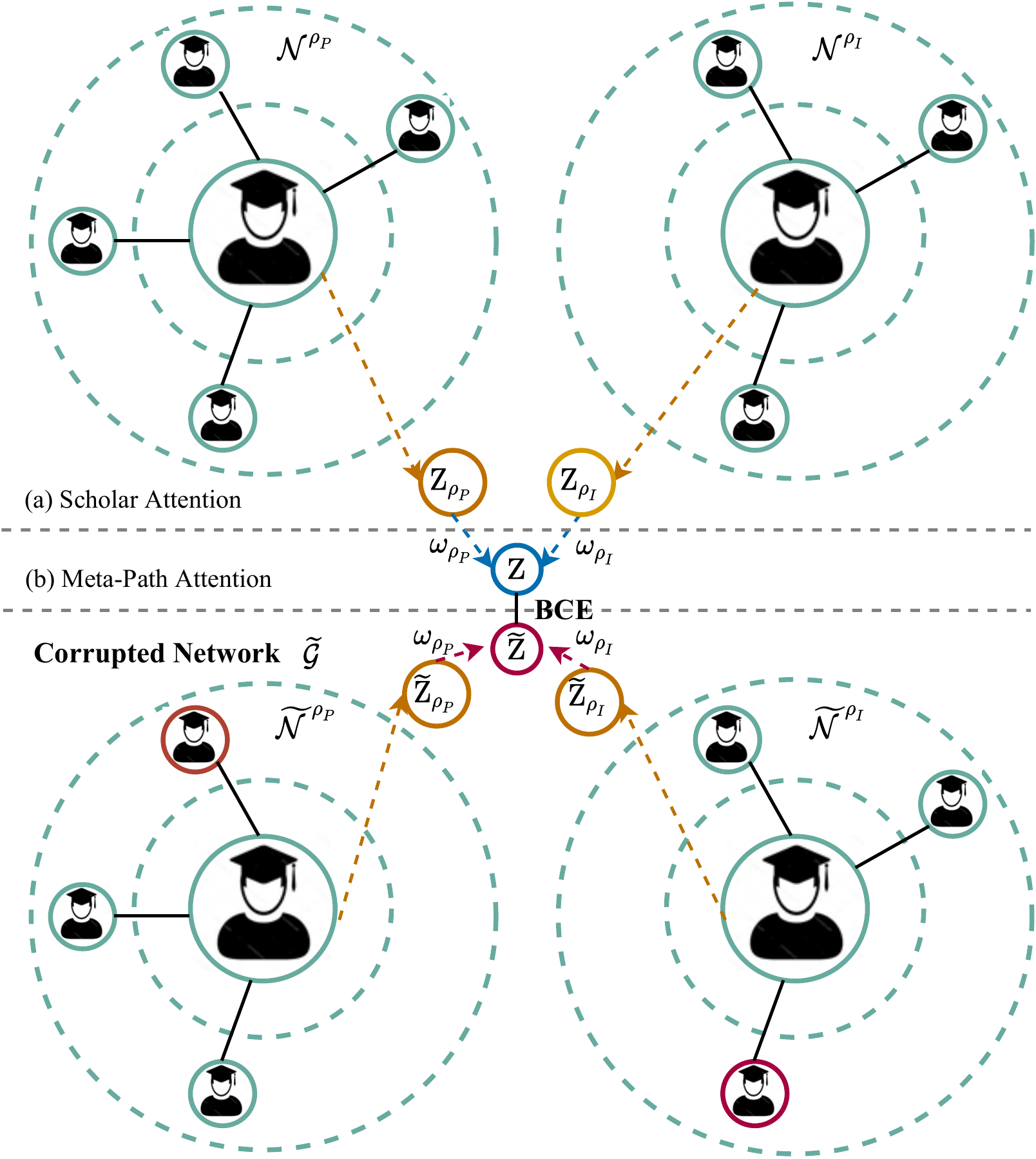}
    \caption{The overall framework of HAI.}
    \label{overall}
\end{figure}

Figure~\ref{overall} illustrates the overall framework of the model HAI. The black lines in the figure are second-order links connected by meta-paths. Two kinds of meta-path play a significant role in calculating scholar attentions. Figure~\ref{overall}(a) shows the process of scholar attention on a specific scholar based on two meta-path neighbors $\mathcal{N}^{\rho_P}$ and $\mathcal{N}^{\rho_I}$. Figure~\ref{overall}(b) shows the process of the meta-path attention which aggregates different kinds of scholar attentions. More about scholar attention, meta-path attention ,and mutual information are detailed in the following sections.
\subsection{Scholar Attention}
Supervisors and collaborators of early-career researchers may influence their choice on which target institutions to apply for. Hence, we present the scholar attention to learn the influence by calculating the attention score through meta-path neighbors and represent the embeddings for each scholar in the HSN.

We leverage Doc2vec~\cite{le2014distributed} to represent the abstracts of each scholar into feature spaces. The projection process can be formulated as follows:
\begin{equation}
    \label{doc2vec}
    H_i = \sum_{k=1}^{N_i} Doc2vec(ab_k)/N_{i},
\end{equation}
where $H_i$ denotes the represented feature of scholar $i$, $N_i$ and $ab_k$ is the number of papers and the $k$th abstract of the paper that scholar $i$ had published, respectively.

Next, we leverage the masked self-attention mechanism~\cite{vaswani2017attention, gat2018graph} to learn the information between collaborators and colleagues. Given a scholar pair $(i,j)$ that are connected with meta-path $\rho$, the attention score $\alpha_{i,j}$ can learn how relevant is the scholar $j$ to the scholar $i$ on the network. Notice that $\alpha_{i,j} \neq \alpha_{j,i}$ because they have different orders when calculating attention scores with masked attention. The attention score of scholar pair $(i,j)$ on meta-path $\rho$ can be formulated as follows:
\begin{equation}
    \label{attScore}
    e_{i,j}^\rho = att_{scholar}(\mathbf{W} H_i,\mathbf{W} H_j, \rho),
\end{equation}
where $att_{scholar}$ denotes the shared attentional mechanism on the deep neural network which performs the scholar attention through meta-path $\rho$, and the weight matrix $\mathbf{W}\in \mathbb{R} ^{D\times D}$ is a shared linear transformation which is applied to every node on meta-path $\rho$.

We only calculate the attention score on the meta-path neighbors, so there comes the masked attention. The masked attention will mask other nodes and only operate on each scholar $j\in \mathcal{N}^\rho_i$, where $\mathcal{N}^\rho_i$ denotes the meta-path neighbors of scholar $i$. In this study, $j$ will be the second-order neighbor of $i$ on meta-path \emph{APA} or \emph{AIA}. To make weight coefficients more clear to compare different scholars, we normalize them by using an activate function softmax:
\begin{equation}
    \label{attSoft}
    \alpha_{i,j}^\rho =\mathit{softmax}_j(e_{i,j}^\rho),
\end{equation}
which can be expanded as:
\begin{equation}
    \label{attSoftexpand}
    \alpha_{i,j}^\rho =\frac{\mathrm{exp}(\sigma (\mathrm{a}^\mathrm{T}[\mathbf{W} H_i\|\mathbf{W} H_j ] ))}{\begin{matrix} \sum_{k\in \mathcal{N}^\rho_i} \mathrm{exp}(\sigma (\mathrm{a}^\mathrm{T}[\mathbf{W} H_i\|\mathbf{W} H_j ] )) \end{matrix}},
\end{equation}
where $\alpha_{i,j}^\rho$ denotes the scholar attention weight coefficient, $\sigma$ is the activate function such as LeakyReLU, $\cdot ^\mathrm{T}$ and $\|$ is the transpose operation and the concatenation operation on matrix. The Figure~\ref{overall} (a) shows the process of scholar attention.

Corresponding the scholar attention weight coefficients, we can obtain the aggregated scholar features of $a_i$ by the operation formulated as follows:
\begin{equation}
    \label{attZ}
    z^\rho_i=\sigma\left(\sum_{j\in \mathcal{N}^\rho_i}\alpha_{i,j}^\rho\mathbf{W} H_j\right),
\end{equation}
where $z^\rho_i$ is the embedding of scholar $a_i$ on meta-path $\rho$. To learn more stable embeddings on the heterogeneous network, we have found expanding Equation~\ref{attZ} to multi-head will benefit attention process a lot. Embedding features of each head that are concatenated results in the final scholar attention feature representation formulated as follows:
\begin{equation}
    \label{attFinal}
    z^\rho_i=\overset{K}{\underset{k=1}{\Vert }}\sigma\left(\sum_{j\in \mathcal{N}^\rho_i}\alpha_{i,j}^{\rho,k}\mathbf{W}^k H_j\right).
\end{equation}
Here, $\alpha_{i,j}^{\rho,k}$ represents the $k$th-head  scholar  attention  weight  coefficient on meta-path $\rho$, and $\mathbf{W}^k$ is the $k$th  weight  matrix. We denote $\mathbf{Z}_{\rho_i}$ as the feature representation on the $i$th meta-path.
\subsection{Meta-Path Attention}
For a specific scholar, we aggregate scholar attention weight coefficients from different meta-paths to one representation space. The meta-path attention will learn the weights of different types of meta-paths, which does not have the same influence on scholar representations. In the constructed heterogeneous network, we denote the meta-path \emph{APA} and \emph{AIA} as $\rho_P$ and $\rho_I$, respectively. Every scholar in the heterogeneous network has two types of semantic information, which significantly differs from that of the homogeneous network. Making full use of the heterogeneity of the network, the meta-path attention automatically learns the weights of different meta-path and aggregates the weights to representation space for the institution recommendation. To learn the different influence of meta-path \emph{APA} and \emph{AIA} on each scholar, we deploy the final nonlinear transformation layer to measure the influence which can be formulated as follows:

\begin{equation}
    \label{attMeta}
    \omega_{\rho_i}=\sigma \left(\frac{1}{\left\lvert \mathcal{V}^a\right\rvert  }\sum_{j\in \mathcal{V}^a}\mathrm{q}^\mathrm{T}\tanh (\mathbf{W}z^{\rho_i}_j+\mathrm{b})\right) ,
\end{equation}
where $\mathrm{b}$ is the bias vector, $\mathrm{q}$ is a scholar attention vector, and $z^{\rho_i}_j$ represents the feature representation of scholar $j$ on the $i$th meta-path $\rho_i$. The higher $\omega_{\rho_i}$ is, the more important meta-path $\rho_i$ is for scholars on the institution recommendation task. We finally get the embeddings of each scholar by aggregate $\omega_{\rho_i}$ and scholar attention feature $\mathbf{Z}_{\rho_i}$:
\begin{equation}
    \label{finalEbed}
    \mathbf{Z}=\sum^{\left\lvert \mathcal{P}\right\rvert }_{i=1}\omega_{\rho_i} \mathbf{Z}_{\rho_i}.
\end{equation}

\subsection{Scholar Mutual Information}
In the institution recommendation task, we do not have any labels about where the scholar will move to. Therefore, we propose a novel unsupervised framework to tackle this problem by importing Deep InfoMax~\cite{hjelm2018learning,velickovic2018deep} into heterogeneous networks.
The original Deep InfoMax algorithm learns the representation for the downstream task by maximizing the mutual information between the input and the output of encoder, while our approach learns the representation for the recommendation by maximizing the mutual information between local features and global features. We set the scholar embedding $\mathbf{Z}$ to be the local features and obtain the global features $s $ by a readout function, which can be formulated as follows:
\begin{equation}
    \label{readout}
    \mathcal{R}(\mathbf{Z})=\frac{1}{\left\lvert \mathcal{V} \right\rvert } \sum_{i = 1}^{\left\lvert \mathcal{V} \right\rvert } \mathrm{z}_i .
\end{equation}
Next, some negative samples are provided by constructing a fake heterogeneous network from the original one. Shuffling the nodes in each type of node, a corrupted function $\mathcal{C} $ is constructed. Subsequently, we get a fake heterogeneous network $\widetilde{\mathcal{G}}= \mathcal{C}(\mathcal{G}) =(\widetilde{\mathcal{V}}  ,\widetilde{\mathcal{E}} ,\widetilde{\mathcal{A}}  ,\widetilde{\mathcal{X}}   ) $. A discriminator is deployed to classify the negative samples, which is a linear binary classification formulated as follows:
\begin{equation}
    \label{discriminator}
    \mathcal{D} (z_i,s)= \sigma (z_i^\mathrm{T}\mathbf{W}s).
\end{equation}
Following the intuitions from Deep infomax, we use a noise-contrastive type objective with a standard binary cross-entropy (BCE) between positive samples and negative samples as the objective
function:
\begin{equation}
    \begin{split}
    \mathcal{L}
        &= \frac{1}{N+M} \Bigg( \sum_{i = 1}^{N} \mathbb{E}_{\left(\mathcal{X},\mathcal{A}\right)  } \left[\log \mathcal{D} (z_i,s)\right] \\
        &+  \sum_{i = 1}^{M} \mathbb{E}_{(\tilde{\mathcal{X}} ,\tilde{\mathcal{A}} ) } \left[\log \left(1-\mathcal{D} \left(\tilde{z_i} ,\tilde{s}\right) \right)  \right]\Bigg).
    \end{split}
\end{equation}
This approach effectively maximizes the mutual information between local features and global features by the Adam optimizer.

\subsection{Model Analysis}
The analysis of the model HAI are described as follows:
\begin{itemize}
    \item HAI handles various types of nodes and edges, which can be trained on the heterogeneous network without labels.
    \item The overall algorithmic complexity of HAI is $O (VC_{att})$,
    where $V$ is the number of nodes, and $C_{att}$ represents the cost of attention operation on the node pairs of meta-path neighbors. $C_{att}=O(KNF)$ where $K$ is the number of attention heads, $N$ is the number of meta-path neighbors, and $F$ is the number of input features. The low cost of computation makes HAI, which is linear to the number of nodes, efficiently capture the information from meta-path by maximizing mutual information and make decisions for the downstream institution recommendation task.
    \item A general challenge of heterogeneous network embedding is the low interpretability (“black box problem”), while the model we proposed has good explainability using the attention mechanism. For a certain scholar in the HSN, HAI can automatically discover the relationships between collaborators and colleagues by calculating the attention score on the list of meta-path neighbors. Meanwhile, which scholar or meta-path is more related to the task can be discovered to explain the results of our model.
\end{itemize}

\section{Experiments}\label{section5}
In this section, we analyze the patterns of scholar career moves and present the results of the experiments to evaluate the efficiency of our model on the institution recommendation task and show the explainability.

\subsection{Preprocessing}
The purpose of preprocessing is to construct an HSN and generate testing set to validate our proposed model. The model is designed for those junior scholars whose academic year is between five and ten years.
\begin{table}[!htbp]
    \centering
    \caption{\label{dataset}Heterogeneous  Scholarly  Network Statistics}
    \resizebox{\linewidth}{!}{
        \begin{tabular}{cccccc}
            \toprule
            \# Scholars & \# Papers             & \# Institutions            & \# Works-With            & \# Writes      \\
            \midrule
            2563 & 3992            & 100            & 2563            & 4276      \\
            \bottomrule
            \end{tabular}
    }
\end{table}
\begin{figure*}[!bp]
    \centering
    \subfigure[HR@K of 64 embedding dimensions]
    {
        \includegraphics[width=1.6in]{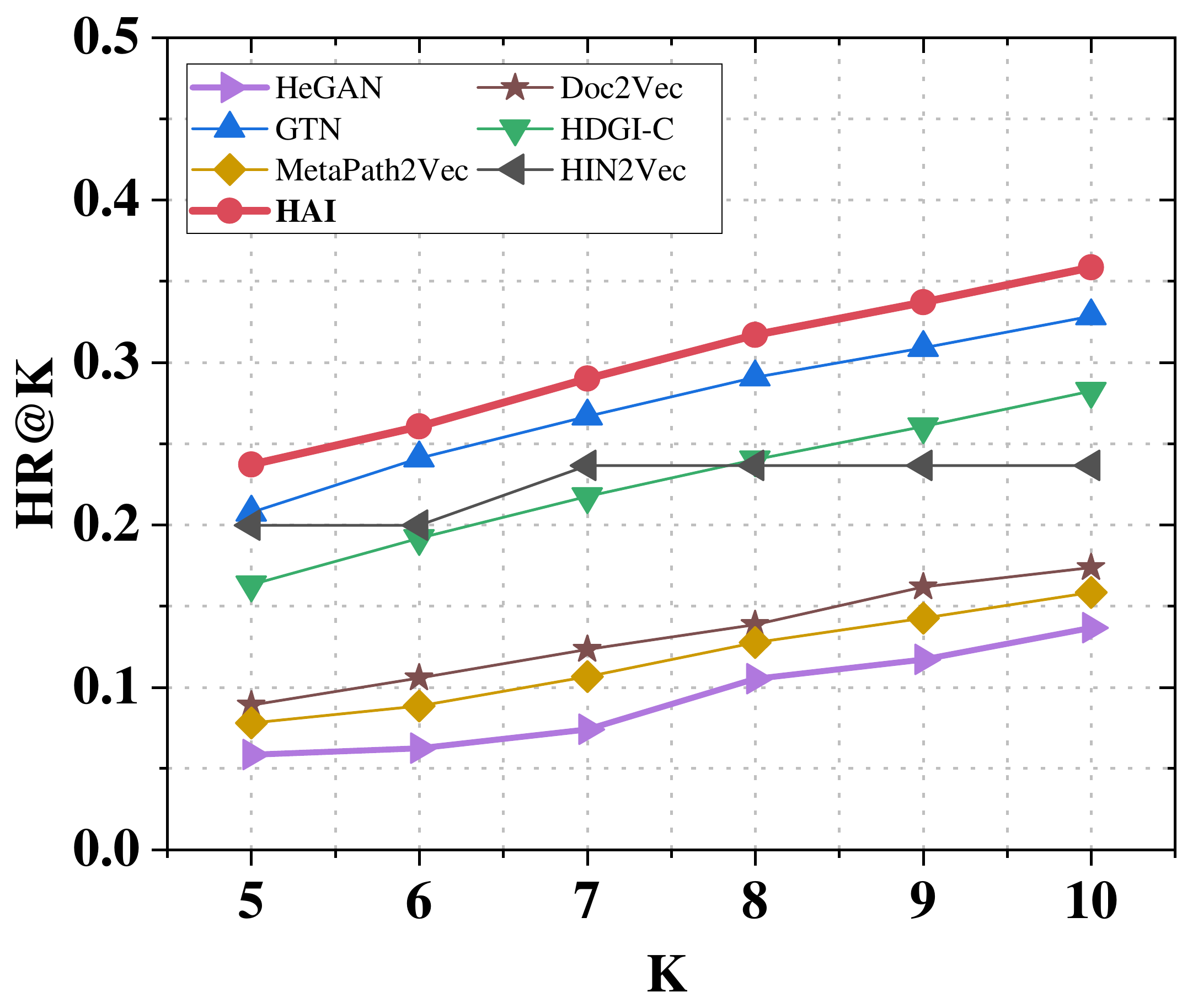}
        \label{fig:first_sub}
    }
    \subfigure[HR@K of 128 embedding dimensions]
    {
        \includegraphics[width=1.6in]{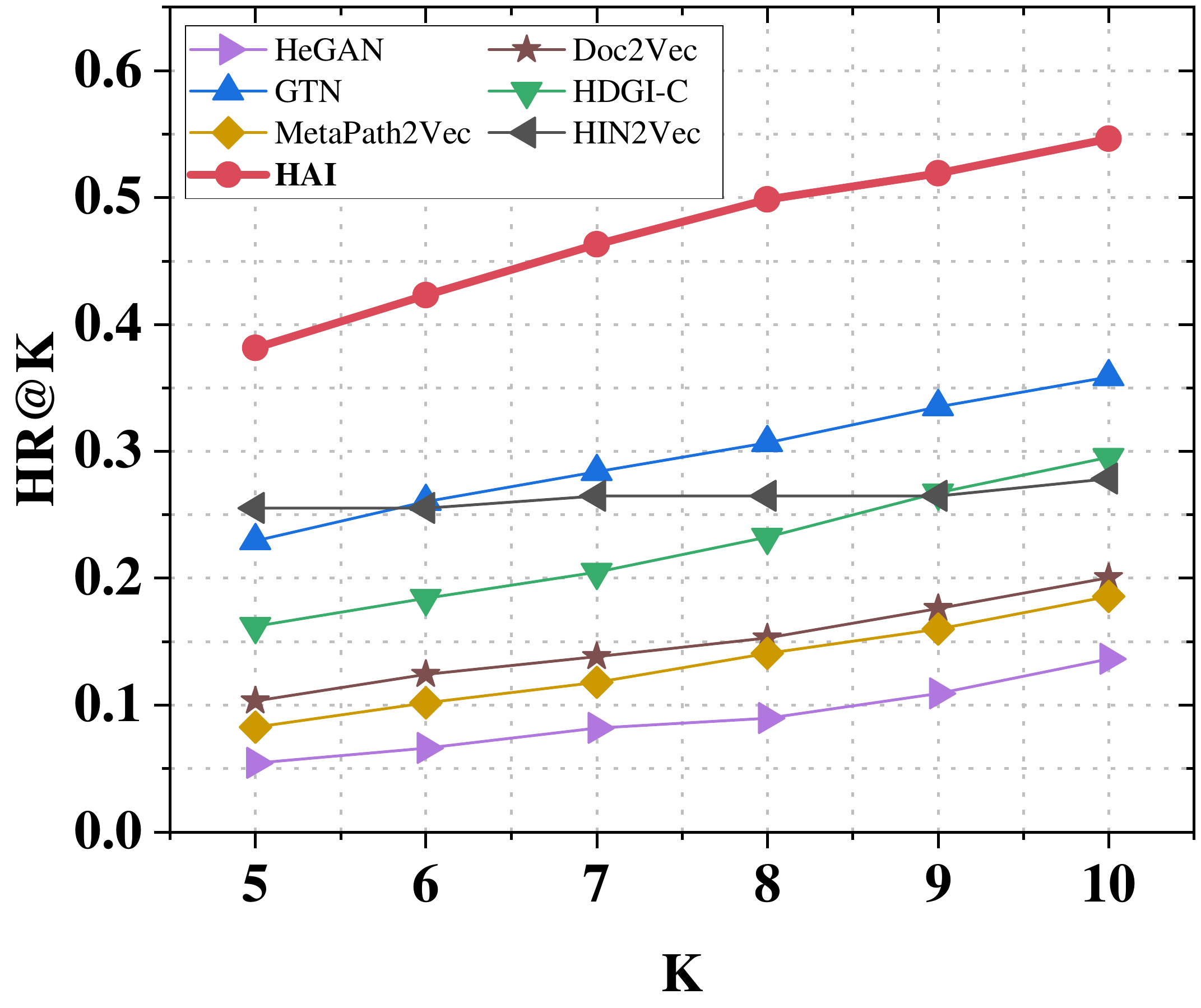}
        \label{fig:second_sub}
    }
    \subfigure[HR@K of 256 embedding dimensions]
    {
        \includegraphics[width=1.6in]{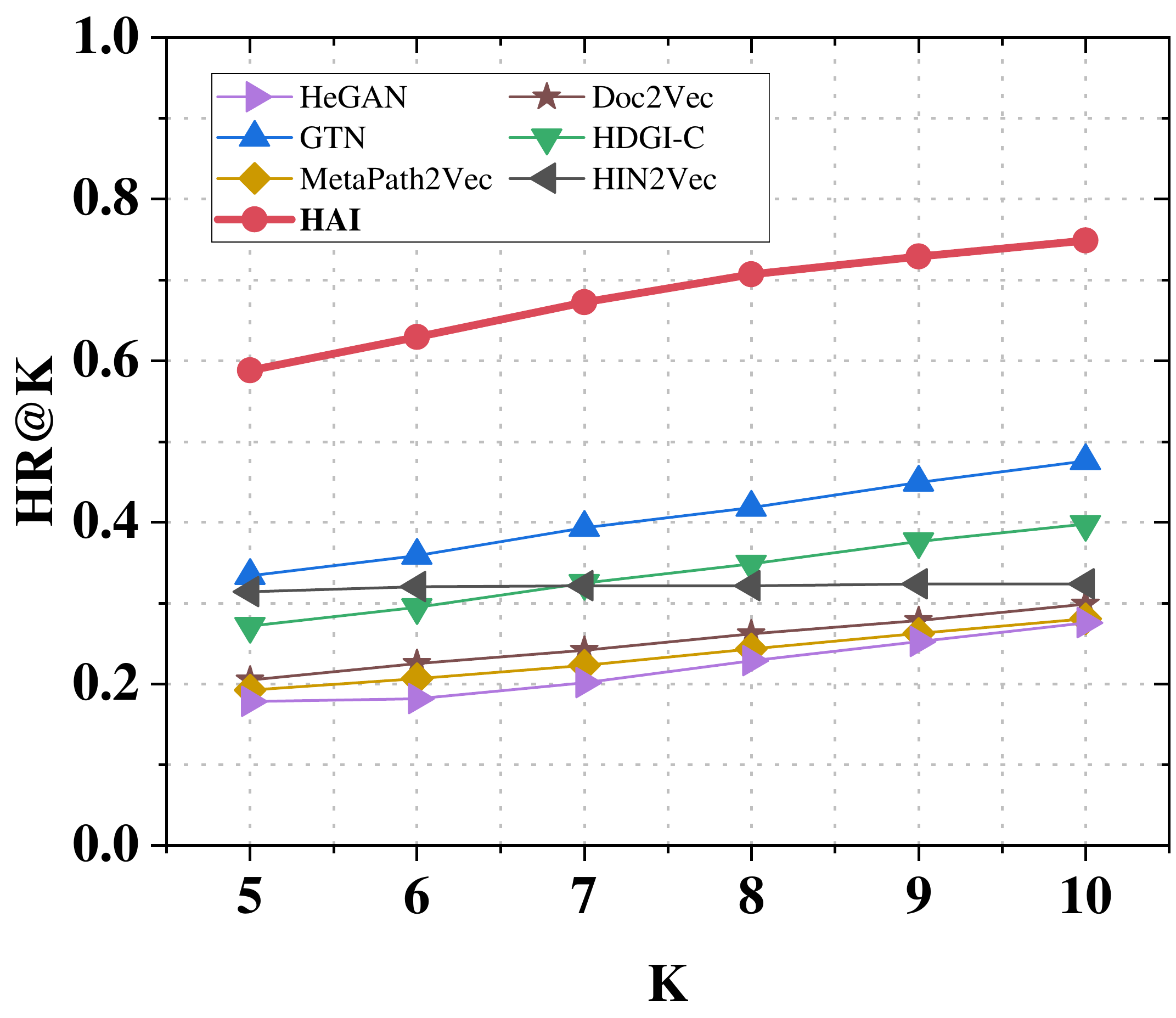}
        \label{fig:third_sub}
    }
    \subfigure[HR@K of 512 embedding dimensions]
    {
        \includegraphics[width=1.6in]{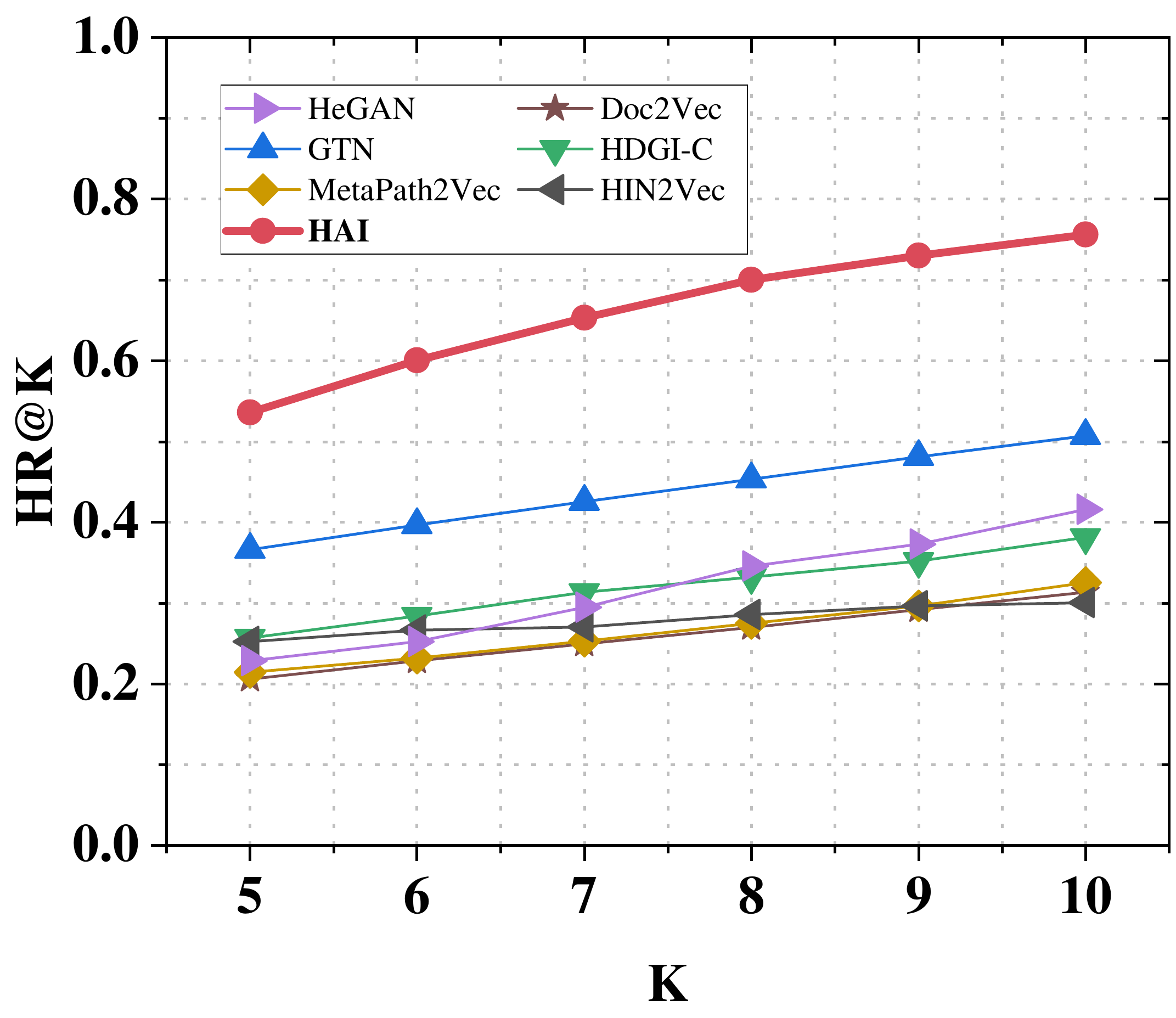}
        \label{fig:fourth_sub}
    }
    \caption{The Hit Ratio at top $K$ of different embedding dimensions.}
    \label{fig:topk}
\end{figure*}
In this study, we construct an HSN with three types of nodes, containing authors, papers, and institutions. Considering nodes in different types have different features, we set the abstract of papers as the features of papers, the abstract of papers that authors have published during 2010 to 2015 as the features of authors, and the features of institutions are encoded to one-hot embeddings especially. If we represent the institution features in the same way of authors and papers, there may be over-smoothing in the representation space because of too many overlapping papers in different institutions.
We construct the HSN on the ASN~\cite{tang2008arnetminer} which is extracted by the AMiner from DBLP, ACM, MAG (Microsoft Academic Graph), and other sources. We collect the top 100 institutions in terms of the number of articles published between 2010 and 2015. Further, we collect the postgraduate students who satisfy the following three conditions:
\begin{enumerate}
    \item academic age ranging from 5 to 10,
    \item have published papers when worked with the top 100 institutions during 2010 to 2015,
    \item staying in the same institution during these 5 years.
\end{enumerate}

Table~\ref{dataset} shows the analysis of HSN, which has 6,655 nodes totally. The nodes consist of 2563 scholars, 3992 papers, and 100 institutions. Additionally, it contains 2563 edges connected with scholars and institutions and 4276 edges connected with scholars and papers.
To evaluate the effectiveness of our model, we extract the institution from where scholars firstly employed during 2015 to 2020 as the testing set. We utilize the meta-path \emph{APA} and \emph{AIA} to perform the experiment.

\subsection{Experimental Settings}
\subsubsection{Baselines}
We evaluate HAI on the institution recommendation task against three state-of-the-art unsupervised heterogeneous network learning models HeGAN~\cite{hu2019adversarial}, GTN~\cite{yun2019graph}, HDGI-C~\cite{ren2019heterogeneous}; two classical unsupervised heterogeneous network learning models MetaPath2Vec~\cite{dong2017metapath2vec}, HIN2Vec~\cite{fu2017hin2vec}; one classical recommendation algorithm Collaborative Filtering~\cite{sarwar2001item}; and one representation learning model Doc2Vec~\cite{le2014distributed}.
\begin{itemize}
    \item \emph{Collaborative Filtering}~\cite{sarwar2001item} is a classical widely deployed recommendation algorithm in the industry.
    \item \emph{Doc2Vec}~\cite{le2014distributed} is a model that represents arbitrary documents to a specific feature space.
    \item \emph{MetaPath2Vec}~\cite{dong2017metapath2vec} is a model for heterogeneous graph embedding, which generates meta-path based on random walks and embeds nodes through the skip-gram algorithm.
    \item \emph{HIN2Vec}~\cite{fu2017hin2vec} is a neural network based model for heterogeneous network representation learning, which utilizes random walk and negative sampling to generate meta-path and represent nodes and meta-paths through neural networks.
    \item \emph{HeGAN}~\cite{hu2019adversarial} is a deep model into adversarial learning on heterogeneous information networks, which is inspired by generative adversarial networks.
    \item \emph{GTN}~\cite{yun2019graph} is a neural network based model for heterogeneous network, which automatically select the best meta-path.
    \item \emph{HDGI-C}~\cite{ren2019heterogeneous} is a model that aggregates mutual information and Graph Convolutional Networks (GCN).
\end{itemize}

\subsubsection{Evaluation Metric}
To quantitatively evaluate the performance of our model on the institution recommendation, we consider two widely used performance metrics in the recommendation system: AUC and Hit Ratio (HR). The metric AUC is one of the popular metrics used in the industry, which stands for “Area Under The ROC Curve”. The ROC curve~\cite{fawcett2006introduction} is plotted with TPR (true positive rate) against the FPR (true positive rate). ROC is a probability curve and AUC represents the degree or measure of separability. It indicates how well the model is able to distinguish between classes: the higher the AUC is, the better the model performs at recommendation. The metric HR~\cite{shi2019deep} is defined in Equation~\ref{hr}:
\begin{equation}
    \label{hr}
    HR=\frac{\# hits}{\# scholars},
\end{equation}
where $\# hits$ is the number of scholars whose ground-truth institution appears in the top $K$ institutions of preference list $Pre$ we recommend. In our experiment, we truncate the ranked list at $K\in [5\%,6\%,7\%,8\%,9\%,10\%]$. We utilize the percentage of the rank list due to the difference in the length of preference list in the different models.

\subsection{Results and Analysis}
\begin{table}[!hbp]
    \centering
    \caption{\label{auc}AUC Comparisons of Different Methods under Different Dimensions and Execution Time}
    \resizebox{\linewidth}{!}{
        \begin{tabular}{cccccc}
            \toprule
            Models & $D$64             & $D$128            & $D$256            & $D$512     & Time       \\
            \midrule
            Doc2Vec      & 0.6209          & 0.6309          & 0.6378          & 0.6412     &   \textbf{3.16m}  \\
            MetaPath2Vec      & 0.6915          & 0.7466          & 0.7601          & 0.7701     &19.26m     \\
            HIN2Vec      & 0.6440          & 0.7377          & 0.7513          & 0.6906     &23.34 m     \\
            HeGAN      & 0.6674          & 0.6677          & 0.6769          & 0.7062  &40.2m        \\
            GTN      & \textbf{0.7353 }         & 0.7581          & 0.7672          & 0.7880      &53.46m    \\
            HDGI-C      & 0.7186          & 0.7212          & 0.7283          & 0.7122     &20.30m     \\
            \textbf{HAI}   & 0.7352          & \textbf{0.8442} & \textbf{0.8798} & \textbf{0.8973} & 13.2m \\
            \bottomrule
            \end{tabular}
    }
\end{table}

Table~\ref{auc} shows the AUC of different methods on the different feature space dimension $D$ and the average running time of training.
The collaborative filtering with AUC 0.7887 outperforms other baselines even though it is not a representation learning method, which shows that collaborative information on the ASN deeply influences the behavior of career moves.
Experiment shows that the change in dimensionality has little effect on the performance of Doc2Vec, and Doc2Vec shows weak performance on the institution recommendation task, which indicates that a single abstract feature of scholars without structural information has no positive effect on the results. MetaPath2Vec and HIN2Vec are random walk based models for embedding heterogeneous graph, which show good performances on recommendation but consuming too much time. These two random walk based models randomly generate millions of meta-paths on our heterogeneous scholarly graph, costing too much time for walking on each node. The generative adversarial based model HeGAN shows a good performance, but shows instability while training. GTN with mutual information performs second only to our model, while performance improves slowly as the dimension rises. Our model outperforms other baselines on most conditions, and the AUC score close to 0.9 indicates that the ground-truth institution is ranked on the top.

Figure~\ref{fig:topk} shows the HR of different dimensions. Only semantic information based model Doc2Vec and only structure information based model MetaPath2Vec both get a poor performance at HR, while our model that aggregates semantic information and structure information outperforms other baselines. Models that utilize the semantic information of the scholar's abstract have significantly higher Hit Ratio values than models that only utilize structural information. The MetaPath2Vec performs good at AUC but average at HR indicates that the model ranks most ground-truth institutions between 20 and 30 percent of the preference list but not ranks the ground-truth institutions high enough to hit top $K$. Although the GTN with mutual information performs similar to our model at 64 feature dimensions, the transformer of GTN consumes lots of time to transform the network while our model only needs about ten iterations to produce good results in a few minutes. Overall, we recommend using our model to represent scholars into 128-dimensional feature space for institution recommendation to achieve a balance between performance and time-consuming.

\subsection{Case Study}

\subsubsection{Attention Behavior}

\begin{figure}[h]
    \centering
    \includegraphics[width=2.5in]{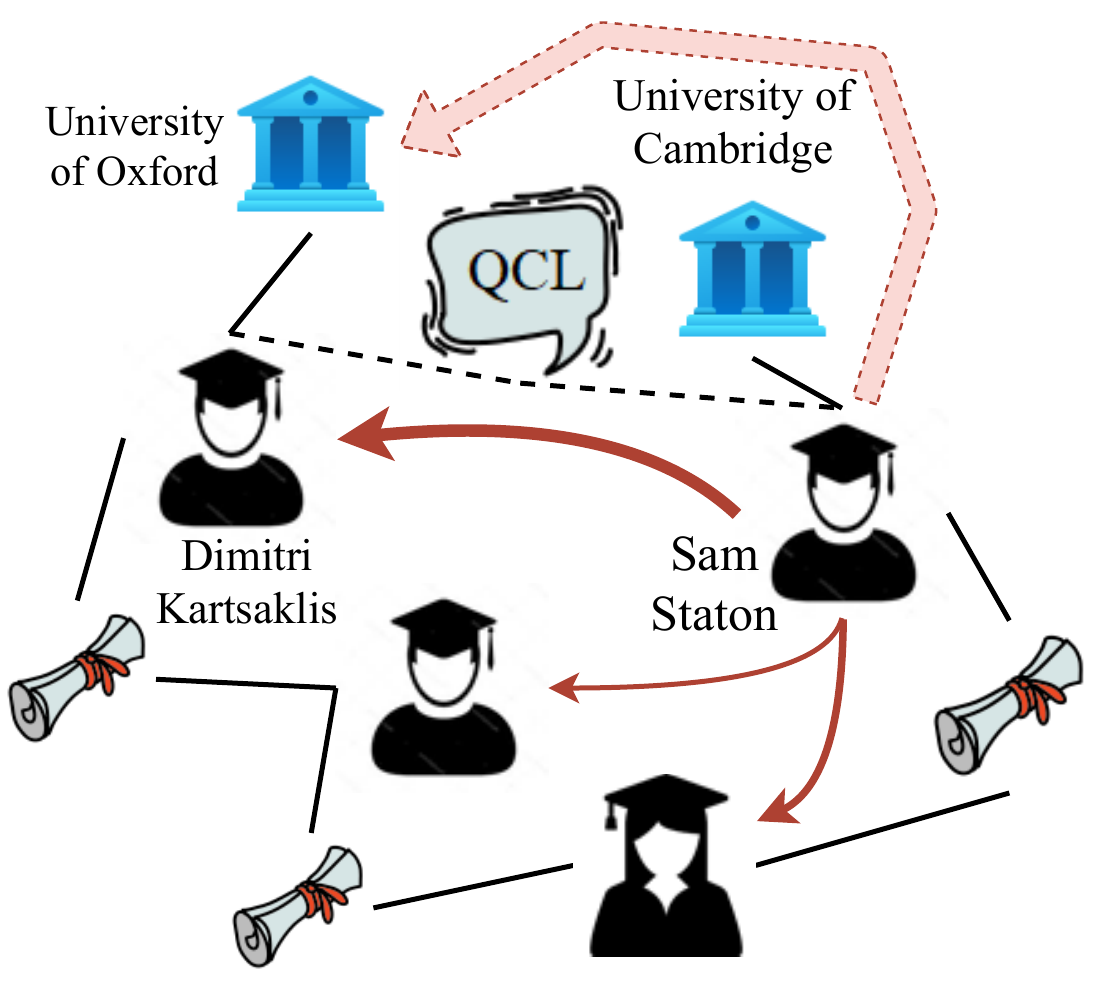}
    \caption{The figure illustrates a case study on how the attention mechanism carries on the meta-path neighbors. The black solid lines represent the edge in the network. The dotted lines indicate the hidden research topic relationships in the network. The red arrows indicate the attention scores and the thickness of the arrow indicates the value of the score. The red arrow with a dashed boundary indicates the career move.}
    \label{casestudy}
\end{figure}
With the case study, we find that even without collaboration information, HAI still can recommend for early-career researchers institutions by research topic. Figure~\ref{casestudy} illustrates the true case of our model recommending institution to the early-career researcher Sam Staton.
Sam Staton and other scholars are on the list of \emph{APA} meta-path neighbors, where attention is carried on. We can see that the scholar attention score from Sam Staton to Dimitri Kartsaklis is significantly higher than others, at the same time Dimitri Kartsaklis is ranked on the top of preference list by our model.
Thus, our model recommends Sam Staton to move to the institution Dimitri Kartsaklis works with. Sam Station moved from University of Cambridge to University of Oxford in March 2015, which matched our recommendation result. Even Sam Station had not collaborated with Dimitri Kartsaklis, the meta-path in scholarly network reveal the similarity between them. And the papers they had published embedded to the node features reveal that they have the same research interest ``Quantum Computing Languages (QCL)''.

\subsubsection{Recommendation Influence}
To quantitatively demonstrate the effectiveness of our recommendations, we analyze the average publication numbers and average citation numbers of Hit scholars and Non-Hit scholars between 2015 and 2020. The scholars are from 100 famous institutions over the world and have similar academic abilities.
We consider those scholars that joined the institution in the Top $K$ of preference list our model recommended as Hit scholars. Similarly, we consider those who did not join the institution in the Top $K$ of preference list as Non-Hit scholars.
\begin{figure}[h]
    \centering
    \includegraphics[width=2.7in]{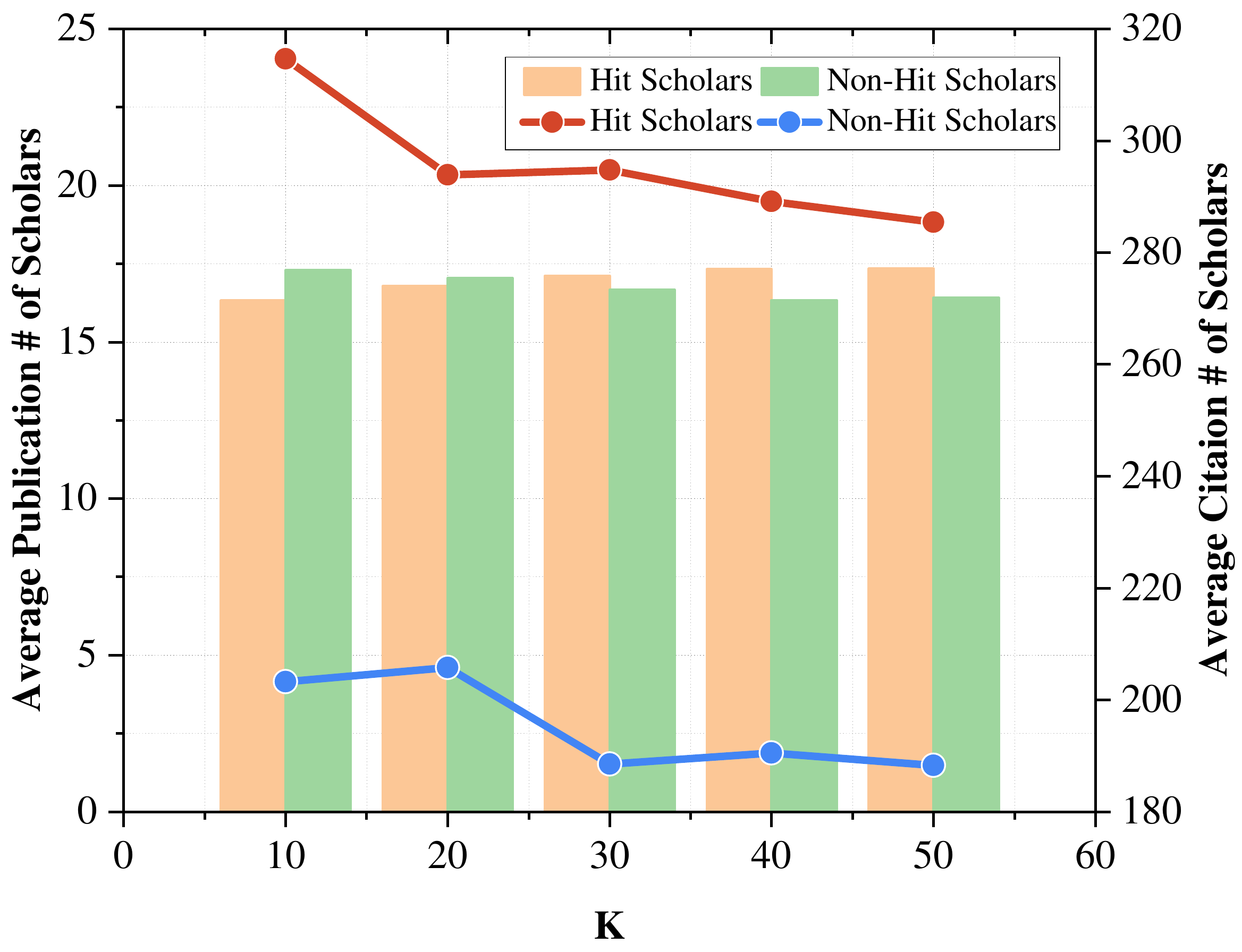}
    \caption{The bar graph illustrates the average publication numbers of Hit scholars and Non-Hit scholars between 2015 and 2020. The line graph illustrates the average citation numbers of Hit scholars and Non-Hit scholars between 2015 and 2020.}
    \label{case2}
\end{figure}
Figure~\ref{case2} shows that the average numbers of citations are significantly greater for Hit scholars than Non-Hit scholars with a small difference in the average numbers of publications. As the institutions that Hit scholars joined getting ranked further down the preference list, the average numbers of citations get lower and lower.
The above results show that Hit scholars have higher average citations than Non-Hit scholars, which means our recommendations are meaningful for early-career scholars.

\section{Conclusion}\label{section6}
In this work, we found that nearly half of the junior scholars had changed academic institutions between 2010 and 2015, with 74\% moved to their collaborator's institution. Based on these findings, we have designed a novel unsupervised learning algorithm HAI for institution recommendation. We applied an attention mechanism to calculate the scholar attention scores on the meta-path neighbors, and we further integrated the abstract representation embedded by Doc2Vec and semantic information of meta-path to represent local features and global features of scholars. 
Finally, we proposed a mutual information based approach to aggregate these features. Experimental results on the ASN demonstrated the effectiveness and efficiency of our approach.
Based on the findings presented in this paper, it is possible to explore the motivation of senior scholars' career moves, as well as learning scholar information on the dynamic scholarly network in the future.

\bibliographystyle{ACM-Reference-Format}
\bibliography{reference.bib} 


\begin{thebibliography}{40}


\ifx \showCODEN    \undefined \def \showCODEN     #1{\unskip}     \fi
\ifx \showDOI      \undefined \def \showDOI       #1{#1}\fi
\ifx \showISBNx    \undefined \def \showISBNx     #1{\unskip}     \fi
\ifx \showISBNxiii \undefined \def \showISBNxiii  #1{\unskip}     \fi
\ifx \showISSN     \undefined \def \showISSN      #1{\unskip}     \fi
\ifx \showLCCN     \undefined \def \showLCCN      #1{\unskip}     \fi
\ifx \shownote     \undefined \def \shownote      #1{#1}          \fi
\ifx \showarticletitle \undefined \def \showarticletitle #1{#1}   \fi
\ifx \showURL      \undefined \def \showURL       {\relax}        \fi
\providecommand\bibfield[2]{#2}
\providecommand\bibinfo[2]{#2}
\providecommand\natexlab[1]{#1}
\providecommand\showeprint[2][]{arXiv:#2}

\bibitem[\protect\citeauthoryear{Baker}{Baker}{2015}]%
        {baker2015social}
\bibfield{author}{\bibinfo{person}{Monya Baker}.}
  \bibinfo{year}{2015}\natexlab{}.
\newblock \showarticletitle{Social media: A network boost}.
\newblock \bibinfo{journal}{\emph{Nature}} \bibinfo{volume}{518},
  \bibinfo{number}{7538} (\bibinfo{year}{2015}), \bibinfo{pages}{263--265}.
\newblock


\bibitem[\protect\citeauthoryear{Dong, Chawla, and Swami}{Dong
  et~al\mbox{.}}{2017}]%
        {dong2017metapath2vec}
\bibfield{author}{\bibinfo{person}{Yuxiao Dong}, \bibinfo{person}{Nitesh~V
  Chawla}, {and} \bibinfo{person}{Ananthram Swami}.}
  \bibinfo{year}{2017}\natexlab{}.
\newblock \showarticletitle{metapath2vec: Scalable representation learning for
  heterogeneous networks}. In \bibinfo{booktitle}{\emph{Proceedings of the 23rd
  ACM SIGKDD international conference on knowledge discovery and data mining}}.
  \bibinfo{pages}{135--144}.
\newblock


\bibitem[\protect\citeauthoryear{Fawcett}{Fawcett}{2006}]%
        {fawcett2006introduction}
\bibfield{author}{\bibinfo{person}{Tom Fawcett}.}
  \bibinfo{year}{2006}\natexlab{}.
\newblock \showarticletitle{An introduction to ROC analysis}.
\newblock \bibinfo{journal}{\emph{Pattern recognition letters}}
  \bibinfo{volume}{27}, \bibinfo{number}{8} (\bibinfo{year}{2006}),
  \bibinfo{pages}{861--874}.
\newblock


\bibitem[\protect\citeauthoryear{Fu, Lee, and Lei}{Fu et~al\mbox{.}}{2017}]%
        {fu2017hin2vec}
\bibfield{author}{\bibinfo{person}{Tao-yang Fu}, \bibinfo{person}{Wang-Chien
  Lee}, {and} \bibinfo{person}{Zhen Lei}.} \bibinfo{year}{2017}\natexlab{}.
\newblock \showarticletitle{Hin2vec: Explore meta-paths in heterogeneous
  information networks for representation learning}. In
  \bibinfo{booktitle}{\emph{Proceedings of the 2017 ACM on Conference on
  Information and Knowledge Management}}. \bibinfo{pages}{1797--1806}.
\newblock


\bibitem[\protect\citeauthoryear{{Guo}, {Xia}, {Zhen}, {Bai}, {Zhang}, {Liu},
  and {Tang}}{{Guo} et~al\mbox{.}}{2020}]%
        {guo2020graduate}
\bibfield{author}{\bibinfo{person}{Teng {Guo}}, \bibinfo{person}{Feng {Xia}},
  \bibinfo{person}{Shihao {Zhen}}, \bibinfo{person}{Xiaomei {Bai}},
  \bibinfo{person}{Dongyu {Zhang}}, \bibinfo{person}{Zitao {Liu}}, {and}
  \bibinfo{person}{Jiliang {Tang}}.} \bibinfo{year}{2020}\natexlab{}.
\newblock \showarticletitle{Graduate Employment Prediction with Bias}. In
  \bibinfo{booktitle}{\emph{Proceedings of the AAAI Conference on Artificial
  Intelligence}}, Vol.~\bibinfo{volume}{34}. \bibinfo{pages}{670--677}.
\newblock


\bibitem[\protect\citeauthoryear{{Hjelm}, {Fedorov}, {Lavoie-Marchildon},
  {Grewal}, {Bachman}, {Trischler}, and {Bengio}}{{Hjelm}
  et~al\mbox{.}}{2018}]%
        {hjelm2018learning}
\bibfield{author}{\bibinfo{person}{R.~Devon {Hjelm}}, \bibinfo{person}{Alex
  {Fedorov}}, \bibinfo{person}{Samuel {Lavoie-Marchildon}},
  \bibinfo{person}{Karan {Grewal}}, \bibinfo{person}{Philip {Bachman}},
  \bibinfo{person}{Adam {Trischler}}, {and} \bibinfo{person}{Yoshua {Bengio}}.}
  \bibinfo{year}{2018}\natexlab{}.
\newblock \showarticletitle{Learning deep representations by mutual information
  estimation and maximization}. In \bibinfo{booktitle}{\emph{International
  Conference on Learning Representations}}.
\newblock


\bibitem[\protect\citeauthoryear{Hou, Ren, Zhang, Kong, Zhang, and Xia}{Hou
  et~al\mbox{.}}{2020}]%
        {hou2020network}
\bibfield{author}{\bibinfo{person}{Mingliang Hou}, \bibinfo{person}{Jing Ren},
  \bibinfo{person}{Da Zhang}, \bibinfo{person}{Xiangjie Kong},
  \bibinfo{person}{Dongyu Zhang}, {and} \bibinfo{person}{Feng Xia}.}
  \bibinfo{year}{2020}\natexlab{}.
\newblock \showarticletitle{Network embedding: Taxonomies, frameworks and
  applications}.
\newblock \bibinfo{journal}{\emph{Computer Science Review}}
  \bibinfo{volume}{38} (\bibinfo{year}{2020}), \bibinfo{pages}{100296}.
\newblock


\bibitem[\protect\citeauthoryear{Hu, Fang, and Shi}{Hu et~al\mbox{.}}{2019}]%
        {hu2019adversarial}
\bibfield{author}{\bibinfo{person}{Binbin Hu}, \bibinfo{person}{Yuan Fang},
  {and} \bibinfo{person}{Chuan Shi}.} \bibinfo{year}{2019}\natexlab{}.
\newblock \showarticletitle{Adversarial learning on heterogeneous information
  networks}. In \bibinfo{booktitle}{\emph{Proceedings of the 25th ACM SIGKDD
  International Conference on Knowledge Discovery \& Data Mining}}.
  \bibinfo{pages}{120--129}.
\newblock


\bibitem[\protect\citeauthoryear{Hu, Shi, Zhao, and Yu}{Hu
  et~al\mbox{.}}{2018}]%
        {hu2018leveraging}
\bibfield{author}{\bibinfo{person}{Binbin Hu}, \bibinfo{person}{Chuan Shi},
  \bibinfo{person}{Wayne~Xin Zhao}, {and} \bibinfo{person}{Philip~S Yu}.}
  \bibinfo{year}{2018}\natexlab{}.
\newblock \showarticletitle{Leveraging meta-path based context for top-n
  recommendation with a neural co-attention model}. In
  \bibinfo{booktitle}{\emph{Proceedings of the 24th ACM SIGKDD International
  Conference on Knowledge Discovery \& Data Mining}}.
  \bibinfo{pages}{1531--1540}.
\newblock


\bibitem[\protect\citeauthoryear{Le and Mikolov}{Le and Mikolov}{2014}]%
        {le2014distributed}
\bibfield{author}{\bibinfo{person}{Quoc Le} {and} \bibinfo{person}{Tomas
  Mikolov}.} \bibinfo{year}{2014}\natexlab{}.
\newblock \showarticletitle{Distributed representations of sentences and
  documents}. In \bibinfo{booktitle}{\emph{International conference on machine
  learning}}. PMLR, \bibinfo{pages}{1188--1196}.
\newblock


\bibitem[\protect\citeauthoryear{{Lu}, {Shi}, {Hu}, and {Liu}}{{Lu}
  et~al\mbox{.}}{2019}]%
        {lu2019relation}
\bibfield{author}{\bibinfo{person}{Yuanfu {Lu}}, \bibinfo{person}{Chuan {Shi}},
  \bibinfo{person}{Linmei {Hu}}, {and} \bibinfo{person}{Zhiyuan {Liu}}.}
  \bibinfo{year}{2019}\natexlab{}.
\newblock \showarticletitle{Relation Structure-Aware Heterogeneous Information
  Network Embedding}. In \bibinfo{booktitle}{\emph{Proceedings of the AAAI
  Conference on Artificial Intelligence}}, Vol.~\bibinfo{volume}{33}.
  \bibinfo{pages}{4456--4463}.
\newblock


\bibitem[\protect\citeauthoryear{{Qu}, {Tang}, and {Han}}{{Qu}
  et~al\mbox{.}}{2018}]%
        {qu2018curriculum}
\bibfield{author}{\bibinfo{person}{Meng {Qu}}, \bibinfo{person}{Jian {Tang}},
  {and} \bibinfo{person}{Jiawei {Han}}.} \bibinfo{year}{2018}\natexlab{}.
\newblock \showarticletitle{Curriculum Learning for Heterogeneous Star Network
  Embedding via Deep Reinforcement Learning}. In
  \bibinfo{booktitle}{\emph{Proceedings of the Eleventh ACM International
  Conference on Web Search and Data Mining}}. \bibinfo{pages}{468--476}.
\newblock


\bibitem[\protect\citeauthoryear{{Ren}, {Wang}, {Wang}, {Yu}, {Hou}, {Lee},
  {Kong}, and {Xia}}{{Ren} et~al\mbox{.}}{2019}]%
        {ren2019api}
\bibfield{author}{\bibinfo{person}{Jing {Ren}}, \bibinfo{person}{Lei {Wang}},
  \bibinfo{person}{Kailai {Wang}}, \bibinfo{person}{Shuo {Yu}},
  \bibinfo{person}{Mingliang {Hou}}, \bibinfo{person}{Ivan {Lee}},
  \bibinfo{person}{Xiangjie {Kong}}, {and} \bibinfo{person}{Feng {Xia}}.}
  \bibinfo{year}{2019}\natexlab{}.
\newblock \showarticletitle{API: An Index for Quantifying a Scholar’s
  Academic Potential}.
\newblock \bibinfo{journal}{\emph{IEEE Access}}  \bibinfo{volume}{7}
  (\bibinfo{year}{2019}), \bibinfo{pages}{178675--178684}.
\newblock


\bibitem[\protect\citeauthoryear{Ren, Liu, Huang, Dai, Bo, and Zhang}{Ren
  et~al\mbox{.}}{2019}]%
        {ren2019heterogeneous}
\bibfield{author}{\bibinfo{person}{Yuxiang Ren}, \bibinfo{person}{Bo Liu},
  \bibinfo{person}{Chao Huang}, \bibinfo{person}{Peng Dai},
  \bibinfo{person}{Liefeng Bo}, {and} \bibinfo{person}{Jiawei Zhang}.}
  \bibinfo{year}{2019}\natexlab{}.
\newblock \showarticletitle{Heterogeneous deep graph infomax}.
\newblock \bibinfo{journal}{\emph{arXiv preprint arXiv:1911.08538}}
  (\bibinfo{year}{2019}).
\newblock


\bibitem[\protect\citeauthoryear{Sarwar, Karypis, Konstan, and Riedl}{Sarwar
  et~al\mbox{.}}{2001}]%
        {sarwar2001item}
\bibfield{author}{\bibinfo{person}{Badrul Sarwar}, \bibinfo{person}{George
  Karypis}, \bibinfo{person}{Joseph Konstan}, {and} \bibinfo{person}{John
  Riedl}.} \bibinfo{year}{2001}\natexlab{}.
\newblock \showarticletitle{Item-based collaborative filtering recommendation
  algorithms}. In \bibinfo{booktitle}{\emph{Proceedings of the 10th
  international conference on World Wide Web}}. \bibinfo{pages}{285--295}.
\newblock


\bibitem[\protect\citeauthoryear{Schafer, Frankowski, Herlocker, and
  Sen}{Schafer et~al\mbox{.}}{2007}]%
        {schafer2007collaborative}
\bibfield{author}{\bibinfo{person}{J~Ben Schafer}, \bibinfo{person}{Dan
  Frankowski}, \bibinfo{person}{Jon Herlocker}, {and} \bibinfo{person}{Shilad
  Sen}.} \bibinfo{year}{2007}\natexlab{}.
\newblock \showarticletitle{Collaborative filtering recommender systems}.
\newblock In \bibinfo{booktitle}{\emph{The adaptive web}}.
  \bibinfo{publisher}{Springer}, \bibinfo{pages}{291--324}.
\newblock


\bibitem[\protect\citeauthoryear{{Shi}, {Han}, {Song}, {Wang}, {Wang}, {Du},
  and {Yu}}{{Shi} et~al\mbox{.}}{2021}]%
        {shi2019deep}
\bibfield{author}{\bibinfo{person}{Chuan {Shi}}, \bibinfo{person}{Xiaotian
  {Han}}, \bibinfo{person}{Li {Song}}, \bibinfo{person}{Xiao {Wang}},
  \bibinfo{person}{Senzhang {Wang}}, \bibinfo{person}{Junping {Du}}, {and}
  \bibinfo{person}{Philip~S. {Yu}}.} \bibinfo{year}{2021}\natexlab{}.
\newblock \showarticletitle{Deep Collaborative Filtering with Multi-Aspect
  Information in Heterogeneous Networks}.
\newblock \bibinfo{journal}{\emph{IEEE Transactions on Knowledge and Data
  Engineering}} \bibinfo{volume}{33}, \bibinfo{number}{4}
  (\bibinfo{year}{2021}), \bibinfo{pages}{1413--1425}.
\newblock


\bibitem[\protect\citeauthoryear{{Shi}, {Hu}, {Zhao}, and {Yu}}{{Shi}
  et~al\mbox{.}}{2019}]%
        {shi2019heterogeneous}
\bibfield{author}{\bibinfo{person}{Chuan {Shi}}, \bibinfo{person}{Binbin {Hu}},
  \bibinfo{person}{Wayne~Xin {Zhao}}, {and} \bibinfo{person}{Philip~S. {Yu}}.}
  \bibinfo{year}{2019}\natexlab{}.
\newblock \showarticletitle{Heterogeneous Information Network Embedding for
  Recommendation}.
\newblock \bibinfo{journal}{\emph{IEEE Transactions on Knowledge and Data
  Engineering}} \bibinfo{volume}{31}, \bibinfo{number}{2}
  (\bibinfo{year}{2019}), \bibinfo{pages}{357--370}.
\newblock


\bibitem[\protect\citeauthoryear{Tang, Zhang, Yao, Li, Zhang, and Su}{Tang
  et~al\mbox{.}}{2008}]%
        {tang2008arnetminer}
\bibfield{author}{\bibinfo{person}{Jie Tang}, \bibinfo{person}{Jing Zhang},
  \bibinfo{person}{Limin Yao}, \bibinfo{person}{Juanzi Li}, \bibinfo{person}{Li
  Zhang}, {and} \bibinfo{person}{Zhong Su}.} \bibinfo{year}{2008}\natexlab{}.
\newblock \showarticletitle{Arnetminer: extraction and mining of academic
  social networks}. In \bibinfo{booktitle}{\emph{Proceedings of the 14th ACM
  SIGKDD international conference on Knowledge discovery and data mining}}.
  \bibinfo{pages}{990--998}.
\newblock


\bibitem[\protect\citeauthoryear{Vaswani, Shazeer, Parmar, Uszkoreit, Jones,
  Gomez, Kaiser, and Polosukhin}{Vaswani et~al\mbox{.}}{2017}]%
        {vaswani2017attention}
\bibfield{author}{\bibinfo{person}{Ashish Vaswani}, \bibinfo{person}{Noam
  Shazeer}, \bibinfo{person}{Niki Parmar}, \bibinfo{person}{Jakob Uszkoreit},
  \bibinfo{person}{Llion Jones}, \bibinfo{person}{Aidan~N Gomez},
  \bibinfo{person}{{\L}ukasz Kaiser}, {and} \bibinfo{person}{Illia
  Polosukhin}.} \bibinfo{year}{2017}\natexlab{}.
\newblock \showarticletitle{Attention is all you need}. In
  \bibinfo{booktitle}{\emph{Proceedings of the 31st International Conference on
  Neural Information Processing Systems}}. \bibinfo{pages}{6000--6010}.
\newblock


\bibitem[\protect\citeauthoryear{{Velickovic}, {Fedus}, {Hamilton}, {Liò},
  {Bengio}, and {Hjelm}}{{Velickovic} et~al\mbox{.}}{2018}]%
        {velickovic2018deep}
\bibfield{author}{\bibinfo{person}{Petar {Velickovic}},
  \bibinfo{person}{William {Fedus}}, \bibinfo{person}{William~L. {Hamilton}},
  \bibinfo{person}{Pietro {Liò}}, \bibinfo{person}{Yoshua {Bengio}}, {and}
  \bibinfo{person}{R.~Devon {Hjelm}}.} \bibinfo{year}{2018}\natexlab{}.
\newblock \showarticletitle{Deep Graph Infomax}. In
  \bibinfo{booktitle}{\emph{International Conference on Learning
  Representations}}.
\newblock


\bibitem[\protect\citeauthoryear{{Veličković}, {Cucurull}, {Casanova},
  {Romero}, {Liò}, and {Bengio}}{{Veličković} et~al\mbox{.}}{2018}]%
        {gat2018graph}
\bibfield{author}{\bibinfo{person}{Petar {Veličković}},
  \bibinfo{person}{Guillem {Cucurull}}, \bibinfo{person}{Arantxa {Casanova}},
  \bibinfo{person}{Adriana {Romero}}, \bibinfo{person}{Pietro {Liò}}, {and}
  \bibinfo{person}{Yoshua {Bengio}}.} \bibinfo{year}{2018}\natexlab{}.
\newblock \showarticletitle{Graph Attention Networks}. In
  \bibinfo{booktitle}{\emph{International Conference on Learning
  Representations}}.
\newblock


\bibitem[\protect\citeauthoryear{Wan, Xia, Kong, Hsu, Huang, and Ma}{Wan
  et~al\mbox{.}}{2020}]%
        {wan2020deep}
\bibfield{author}{\bibinfo{person}{Liangtian Wan}, \bibinfo{person}{Feng Xia},
  \bibinfo{person}{Xiangjie Kong}, \bibinfo{person}{Ching-Hsien Hsu},
  \bibinfo{person}{Runhe Huang}, {and} \bibinfo{person}{Jianhua Ma}.}
  \bibinfo{year}{2020}\natexlab{}.
\newblock \showarticletitle{Deep Matrix Factorization for Trust-Aware
  Recommendation in Social Networks}.
\newblock \bibinfo{journal}{\emph{IEEE Transactions on Network Science and
  Engineering}} \bibinfo{volume}{8}, \bibinfo{number}{1}
  (\bibinfo{year}{2020}), \bibinfo{pages}{511--528}.
\newblock


\bibitem[\protect\citeauthoryear{Wang, Liu, Tang, Tuarob, Xia, Gong, and
  King}{Wang et~al\mbox{.}}{2020}]%
        {wang2020attributed}
\bibfield{author}{\bibinfo{person}{Wei Wang}, \bibinfo{person}{Jiaying Liu},
  \bibinfo{person}{Tao Tang}, \bibinfo{person}{Suppawong Tuarob},
  \bibinfo{person}{Feng Xia}, \bibinfo{person}{Zhiguo Gong}, {and}
  \bibinfo{person}{Irwin King}.} \bibinfo{year}{2020}\natexlab{}.
\newblock \showarticletitle{Attributed collaboration network embedding for
  academic relationship mining}.
\newblock \bibinfo{journal}{\emph{ACM Transactions on the Web (TWEB)}}
  \bibinfo{volume}{15}, \bibinfo{number}{1} (\bibinfo{year}{2020}),
  \bibinfo{pages}{1--20}.
\newblock


\bibitem[\protect\citeauthoryear{{Wang}, {Tang}, {Xia}, {Gong}, {Chen}, and
  {Liu}}{{Wang} et~al\mbox{.}}{2020}]%
        {wang2020collaborative}
\bibfield{author}{\bibinfo{person}{Wei {Wang}}, \bibinfo{person}{Tao {Tang}},
  \bibinfo{person}{Feng {Xia}}, \bibinfo{person}{Zhiguo {Gong}},
  \bibinfo{person}{Zhikui {Chen}}, {and} \bibinfo{person}{Huan {Liu}}.}
  \bibinfo{year}{2020}\natexlab{}.
\newblock \showarticletitle{Collaborative Filtering with Network Representation
  Learning for Citation Recommendation}.
\newblock \bibinfo{journal}{\emph{IEEE Transactions on Big Data}}
  (\bibinfo{year}{2020}), \bibinfo{pages}{1--1}.
\newblock


\bibitem[\protect\citeauthoryear{{Wang}, {Xia}, {Wu}, {Gong}, {Tong}, and
  {Davison}}{{Wang} et~al\mbox{.}}{2021}]%
        {wang2021scholar2vec}
\bibfield{author}{\bibinfo{person}{Wei {Wang}}, \bibinfo{person}{Feng {Xia}},
  \bibinfo{person}{Jian {Wu}}, \bibinfo{person}{Zhiguo {Gong}},
  \bibinfo{person}{Hanghang {Tong}}, {and} \bibinfo{person}{Brian~D.
  {Davison}}.} \bibinfo{year}{2021}\natexlab{}.
\newblock \showarticletitle{Scholar2vec: Vector Representation of Scholars for
  Lifetime Collaborator Prediction}.
\newblock \bibinfo{journal}{\emph{ACM Transactions on Knowledge Discovery From
  Data}} \bibinfo{volume}{15}, \bibinfo{number}{3} (\bibinfo{year}{2021}),
  \bibinfo{pages}{40}.
\newblock


\bibitem[\protect\citeauthoryear{Wang, Ji, Shi, Wang, Ye, Cui, and Yu}{Wang
  et~al\mbox{.}}{2019}]%
        {wang2019heterogeneous}
\bibfield{author}{\bibinfo{person}{Xiao Wang}, \bibinfo{person}{Houye Ji},
  \bibinfo{person}{Chuan Shi}, \bibinfo{person}{Bai Wang},
  \bibinfo{person}{Yanfang Ye}, \bibinfo{person}{Peng Cui}, {and}
  \bibinfo{person}{Philip~S Yu}.} \bibinfo{year}{2019}\natexlab{}.
\newblock \showarticletitle{Heterogeneous graph attention network}. In
  \bibinfo{booktitle}{\emph{The World Wide Web Conference}}.
  \bibinfo{pages}{2022--2032}.
\newblock


\bibitem[\protect\citeauthoryear{Wang and Han}{Wang and Han}{2021}]%
        {wang2021attractive}
\bibfield{author}{\bibinfo{person}{Yakun Wang} {and} \bibinfo{person}{Xiaodong
  Han}.} \bibinfo{year}{2021}\natexlab{}.
\newblock \showarticletitle{Attractive community detection in academic social
  network}.
\newblock \bibinfo{journal}{\emph{Journal of Computational Science}}
  \bibinfo{volume}{51} (\bibinfo{year}{2021}), \bibinfo{pages}{101331}.
\newblock


\bibitem[\protect\citeauthoryear{{Wang}, {Liu}, {Du}, {Wu}, and {Zhang}}{{Wang}
  et~al\mbox{.}}{2019}]%
        {wang2019unified}
\bibfield{author}{\bibinfo{person}{Zekai {Wang}}, \bibinfo{person}{Hongzhi
  {Liu}}, \bibinfo{person}{Yingpeng {Du}}, \bibinfo{person}{Zhonghai {Wu}},
  {and} \bibinfo{person}{Xing {Zhang}}.} \bibinfo{year}{2019}\natexlab{}.
\newblock \showarticletitle{Unified Embedding Model over Heterogeneous
  Information Network for Personalized Recommendation.}. In
  \bibinfo{booktitle}{\emph{Proceedings of the Twenty-Eighth International
  Joint Conference on Artificial Intelligence}}. \bibinfo{pages}{3813--3819}.
\newblock


\bibitem[\protect\citeauthoryear{Wiechetek, Phusavat, and Pastuszak}{Wiechetek
  et~al\mbox{.}}{2020}]%
        {wiechetek2020analytical}
\bibfield{author}{\bibinfo{person}{Lukasz Wiechetek}, \bibinfo{person}{Kongkiti
  Phusavat}, {and} \bibinfo{person}{Zbigniew Pastuszak}.}
  \bibinfo{year}{2020}\natexlab{}.
\newblock \showarticletitle{An analytical system for evaluating academia units
  based on metrics provided by academic social network}.
\newblock \bibinfo{journal}{\emph{Expert Systems with Applications}}
  \bibinfo{volume}{159} (\bibinfo{year}{2020}), \bibinfo{pages}{113608}.
\newblock


\bibitem[\protect\citeauthoryear{Woolston}{Woolston}{2017}]%
        {woolston2017graduate}
\bibfield{author}{\bibinfo{person}{Chris Woolston}.}
  \bibinfo{year}{2017}\natexlab{}.
\newblock \showarticletitle{Graduate survey: A love--hurt relationship}.
\newblock \bibinfo{journal}{\emph{Nature}} \bibinfo{volume}{550},
  \bibinfo{number}{7677} (\bibinfo{year}{2017}), \bibinfo{pages}{549--552}.
\newblock


\bibitem[\protect\citeauthoryear{{Xia}, {Liu}, {Nie}, {Fu}, {Wan}, and
  {Kong}}{{Xia} et~al\mbox{.}}{2020}]%
        {xia2020random}
\bibfield{author}{\bibinfo{person}{Feng {Xia}}, \bibinfo{person}{Jiaying
  {Liu}}, \bibinfo{person}{Hansong {Nie}}, \bibinfo{person}{Yonghao {Fu}},
  \bibinfo{person}{Liangtian {Wan}}, {and} \bibinfo{person}{Xiangjie {Kong}}.}
  \bibinfo{year}{2020}\natexlab{}.
\newblock \showarticletitle{Random Walks: A Review of Algorithms and
  Applications}.
\newblock \bibinfo{journal}{\emph{IEEE Transactions on Emerging Topics in
  Computational Intelligence}} \bibinfo{volume}{4}, \bibinfo{number}{2}
  (\bibinfo{year}{2020}), \bibinfo{pages}{95--107}.
\newblock


\bibitem[\protect\citeauthoryear{{Xia}, {Sun}, {Yu}, {Aziz}, {Wan}, {Pan}, and
  {Liu}}{{Xia} et~al\mbox{.}}{2021}]%
        {xia2021graph}
\bibfield{author}{\bibinfo{person}{Feng {Xia}}, \bibinfo{person}{Ke {Sun}},
  \bibinfo{person}{Shuo {Yu}}, \bibinfo{person}{Abdul {Aziz}},
  \bibinfo{person}{Liangtian {Wan}}, \bibinfo{person}{Shirui {Pan}}, {and}
  \bibinfo{person}{Huan {Liu}}.} \bibinfo{year}{2021}\natexlab{}.
\newblock \showarticletitle{Graph Learning: A Survey}.
\newblock \bibinfo{journal}{\emph{IEEE Transactions on Artificial
  Intelligence}} \bibinfo{volume}{2}, \bibinfo{number}{2}
  (\bibinfo{year}{2021}), \bibinfo{pages}{109--127}.
\newblock


\bibitem[\protect\citeauthoryear{Xia, Wang, Bekele, and Liu}{Xia
  et~al\mbox{.}}{2017}]%
        {xia2017big}
\bibfield{author}{\bibinfo{person}{Feng Xia}, \bibinfo{person}{Wei Wang},
  \bibinfo{person}{Teshome~Megersa Bekele}, {and} \bibinfo{person}{Huan Liu}.}
  \bibinfo{year}{2017}\natexlab{}.
\newblock \showarticletitle{Big scholarly data: A survey}.
\newblock \bibinfo{journal}{\emph{IEEE Transactions on Big Data}}
  \bibinfo{volume}{3}, \bibinfo{number}{1} (\bibinfo{year}{2017}),
  \bibinfo{pages}{18--35}.
\newblock


\bibitem[\protect\citeauthoryear{{Ying}, {He}, {Chen}, {Eksombatchai},
  {Hamilton}, and {Leskovec}}{{Ying} et~al\mbox{.}}{2018}]%
        {ying2018graph}
\bibfield{author}{\bibinfo{person}{Rex {Ying}}, \bibinfo{person}{Ruining {He}},
  \bibinfo{person}{Kaifeng {Chen}}, \bibinfo{person}{Pong {Eksombatchai}},
  \bibinfo{person}{William~L. {Hamilton}}, {and} \bibinfo{person}{Jure
  {Leskovec}}.} \bibinfo{year}{2018}\natexlab{}.
\newblock \showarticletitle{Graph Convolutional Neural Networks for Web-Scale
  Recommender Systems}. In \bibinfo{booktitle}{\emph{Proceedings of the 24th
  ACM SIGKDD International Conference on Knowledge Discovery \&amp; Data
  Mining}}. \bibinfo{pages}{974--983}.
\newblock


\bibitem[\protect\citeauthoryear{{Yu}, {Liu}, {Xia}, {Wei}, and {Tong}}{{Yu}
  et~al\mbox{.}}{2021}]%
        {yu2021how}
\bibfield{author}{\bibinfo{person}{Shuo {Yu}}, \bibinfo{person}{Jiaying {Liu}},
  \bibinfo{person}{Feng {Xia}}, \bibinfo{person}{Haoran {Wei}}, {and}
  \bibinfo{person}{Hanghang {Tong}}.} \bibinfo{year}{2021}\natexlab{}.
\newblock \showarticletitle{How to optimize an academic team when the outlier
  member is leaving}.
\newblock \bibinfo{journal}{\emph{IEEE Intelligent Systems}}
  (\bibinfo{year}{2021}), \bibinfo{pages}{1--1}.
\newblock


\bibitem[\protect\citeauthoryear{{Yun}, {Jeong}, {Kim}, {Kang}, and
  {Kim}}{{Yun} et~al\mbox{.}}{2019}]%
        {yun2019graph}
\bibfield{author}{\bibinfo{person}{Seongjun {Yun}}, \bibinfo{person}{Minbyul
  {Jeong}}, \bibinfo{person}{Raehyun {Kim}}, \bibinfo{person}{Jaewoo {Kang}},
  {and} \bibinfo{person}{Hyunwoo~J. {Kim}}.} \bibinfo{year}{2019}\natexlab{}.
\newblock \showarticletitle{Graph Transformer Networks}. In
  \bibinfo{booktitle}{\emph{33rd Annual Conference on Neural Information
  Processing Systems, NeurIPS 2019}}, Vol.~\bibinfo{volume}{32}.
  \bibinfo{pages}{11960--11970}.
\newblock


\bibitem[\protect\citeauthoryear{Zhang, Song, Huang, Swami, and Chawla}{Zhang
  et~al\mbox{.}}{2019}]%
        {zhang2019heterogeneous}
\bibfield{author}{\bibinfo{person}{Chuxu Zhang}, \bibinfo{person}{Dongjin
  Song}, \bibinfo{person}{Chao Huang}, \bibinfo{person}{Ananthram Swami}, {and}
  \bibinfo{person}{Nitesh~V Chawla}.} \bibinfo{year}{2019}\natexlab{}.
\newblock \showarticletitle{Heterogeneous graph neural network}. In
  \bibinfo{booktitle}{\emph{Proceedings of the 25th ACM SIGKDD International
  Conference on Knowledge Discovery \& Data Mining}}.
  \bibinfo{pages}{793--803}.
\newblock


\bibitem[\protect\citeauthoryear{Zhao, Yao, Li, Song, and Lee}{Zhao
  et~al\mbox{.}}{2017}]%
        {zhao2017meta}
\bibfield{author}{\bibinfo{person}{Huan Zhao}, \bibinfo{person}{Quanming Yao},
  \bibinfo{person}{Jianda Li}, \bibinfo{person}{Yangqiu Song}, {and}
  \bibinfo{person}{Dik~Lun Lee}.} \bibinfo{year}{2017}\natexlab{}.
\newblock \showarticletitle{Meta-graph based recommendation fusion over
  heterogeneous information networks}. In \bibinfo{booktitle}{\emph{Proceedings
  of the 23rd ACM SIGKDD international conference on knowledge discovery and
  data mining}}. \bibinfo{pages}{635--644}.
\newblock


\bibitem[\protect\citeauthoryear{{Zhao}, {Zhang}, {Zhou}, {Li}, {Gong}, and
  {Wang}}{{Zhao} et~al\mbox{.}}{2020}]%
        {zhao2020hetnerec}
\bibfield{author}{\bibinfo{person}{Zhongying {Zhao}}, \bibinfo{person}{Xuejian
  {Zhang}}, \bibinfo{person}{Hui {Zhou}}, \bibinfo{person}{Chao {Li}},
  \bibinfo{person}{Maoguo {Gong}}, {and} \bibinfo{person}{Yongqing {Wang}}.}
  \bibinfo{year}{2020}\natexlab{}.
\newblock \showarticletitle{HetNERec: Heterogeneous network embedding based
  recommendation}.
\newblock \bibinfo{journal}{\emph{Knowledge Based Systems}}
  \bibinfo{volume}{204} (\bibinfo{year}{2020}), \bibinfo{pages}{106218}.
\newblock


\end{thebibliography}

\end{document}